\documentclass{article}
\usepackage[preprint]{log_2022}           

\usepackage{booktabs}            % professional-quality tables
\usepackage{multirow}            % tabular cells spanning multiple rows
\usepackage{amsfonts}            % blackboard math symbols
\usepackage{graphicx}            % figures
\usepackage{duckuments}          % sample images
\usepackage{textcomp}
\usepackage{stackengine}
\usepackage{float}

\usepackage[numbers,compress,sort]{natbib}
\title{An Improved Metric and Benchmark for Assessing the Performance of Virtual Screening Models}

\author[M. Brocidiacono et al.]{%
Michael Brocidiacono\\
\institute{Univerity of North Carolina at Chapel Hill}\\
\email{mixarcid@unc.edu}\And
Konstantin I. Popov\\
\institute{Univerity of North Carolina at Chapel Hill}\\
\email{kpopov@unc.edu}\And
Alexander Tropsha\\
\institute{Univerity of North Carolina at Chapel Hill}\\
\email{alex\_tropsha@unc.edu}
}

\newcommand{\EF}[1]{{EF\textsubscript{#1}}}
\newcommand{\EFB}[1]{{$\text{EF}^B_{\text{#1}}$}}
\begin{document}

\maketitle

\begin{abstract}

Structure-based virtual screening (SBVS) is a key workflow in computational drug discovery. SBVS models are assessed by measuring the enrichment of known active molecules over decoys in retrospective screens. However, the standard formula for enrichment cannot estimate model performance on very large libraries. Additionally, current screening benchmarks cannot easily be used with machine learning (ML) models due to data leakage. We propose an improved formula for calculating VS enrichment and introduce the BayesBind benchmarking set composed of protein targets that are structurally dissimilar to those in the BigBind training set. We assess current models on this benchmark and find that none perform appreciably better than a KNN baseline. We publicly release the BayesBind benchmark at \url{https://github.com/molecularmodelinglab/bigbind}.

\end{abstract}

\section{Introduction}

Structure-based virtual screening (SBVS) aims to identify compounds that bind to a protein target. Given the 3D structure of the target's binding site, a large library of compounds is scored according to their predicted ability to bind to a protein target, and the top-scoring molecules are selected for experimental validation. 

Running a prospective virtual screen is time-consuming and expensive. As such, we desire to validate SBVS models virtually before using them in production. An ideal \textit{in silico} benchmark would allow us to 1) select the best model out of a set of candidates and 2) give us some indication of how successful this model will perform in a real-life setting.

The current strategy to validate SBVS models is to create separate benchmarking sets for many protein targets. For each target, compounds that are known to bind to the protein (``actives'') are combined with either computationally identified decoys (DUD-E \cite{mysinger_directory_2012} and CASF \cite{su_comparative_2019}) or experimentally validated inactives (LIT-PCBA \cite{tran-nguyen_lit-pcba_2020}). Models are then assessed according to their ability to discriminate actives from inactives. While traditional classification metrics such as ROC-AUC can be used to assess a model's classification performance, they are not the primary metrics we care about \cite{wellnitz_one_2023}. Instead, we are mainly concerned with whether or not the \textit{top-scoring} molecules are active or inactive.

The simplest and most interpretable early enrichment metric is the enrichment factor (EF), parameterized by a selection fraction $\chi$ (e.g. 1\%). \EF{$\chi$} is defined to be the fraction of actives selected in the top $\chi$ of the compounds divided by the overall fraction of actives in the set. This metric is easily interpreted as the success rate of the model relative to the expected success rate of random selection.

A fundamental issue with the enrichment factor is that the maximum value achievable is the ratio of inactive to active compounds in the set. For DUD-E, the average decoy to active ratio over all the targets is 61. Libraries in real-life screens, however, have much higher inactive-to-active ratios; therefore, models must achieve higher enrichments to be useful. Consider that high-throughput screens (HTS) regularly test hundreds of thousands of random compounds against protein targets, while virtual screens typically suggest hundreds for experimental validation. Thus, for virtual screens to produce results similar to HTS, models must achieve enrichments of around 1,000 (for sufficiently low $\chi$). For the EF formula to measure such high enrichments, we would need benchmarks with extremely high inactive-to-active ratios. Such a benchmark would need to be very large, thus making it prohibitively time-consuming to test models on it. This benchmark would additionally pose problems with using decoys as inactives; with an enormous number of decoys, many of them would likely bind to the protein in real life simply by random chance.

In this work, we propose a new way of calculating enrichments in VS benchmarks. Our improved formula is just as simple to calculate as the original EF and does not suffer from the difficulties outlined above. Notably, the new formula does not assume that decoys in the set are truly inactive. We denote this new metric the Bayes enrichment factor, or \EFB{}, to distinguish it from the traditional formula.

Additionally, we created a new SBVS benchmark, BayesBind, especially for use with machine learning (ML) models. It is currently difficult to evaluate ML models on current VS benchmarking sets due to issues of data leakage. Properly splitting protein-ligand activity datasets is difficult and it is very easy to achieve good performance due to the similarity between the train and test sets \cite{yang_predicting_2020, francoeur_three-dimensional_2020, volkov_frustration_2022}. The targets in the BayesBind benchmark are taken from the validation and test sets of the BigBind dataset \cite{brocidiacono_bigbind_2023}, allowing easy validation of any model that was trained on the BigBind training set. Even with the rigorous splitting used by BigBind, we removed several targets from the set for which a K-nearest-neighbor (KNN) model did suspiciously well on.

\section{The Bayes enrichment factor}

Suppose we are running a virtual screen on $N$ molecules for a particular protein target. Let $A$ be the event that a particular molecule binds to the target, and let $S$ be the score of that molecule given by our SBVS program. $\chi$ is our selection fraction, so we choose the cutoff score $S_\chi$ such that $P(S>S_\chi) = \chi$. We select all molecules with scores above $S_\chi$. We define the ``true'' enrichment of our model at this $\chi$ to be:
\begin{equation}
    \text{EF}^T_\chi = \frac{P(A|S > S_\chi)}{P(A)} 
\end{equation}

 The usual formula for the enrichment factor \EF{$\chi$}  is an estimate of this true enrichment, where, for example, $P(A|S > S_\chi)$ is estimated by the empirical fraction of actives in our chosen set.
 
 Using Bayes' Theorem, it is easy to show that $\text{EF}^T_\chi$ is also equal to $\frac{P(S > S_\chi | A)}{P(S > S_\chi)}$. This ratio can easily be estimated by separately scoring a set of active molecules and a set of \text{random} compounds and observing how their score distributions differ. Thus we define our new metric, the Bayes enrichment factor, to be:

\begin{equation}
    \text{EF}^B_\chi = \frac{\text{Fraction of actives whose score is above $S_\chi$}}{\text{Fraction of random molecules whose score is above $S_\chi$}}
\end{equation}

The \EFB{} has several advantages. Most notably, it requires only random compounds (from the same chemical space as the actives) instead of inactives. This is a major improvement considering how much work has gone into producing decoys that are hypothesized not to bind \cite{mysinger_directory_2012}. This both eliminates a potential source of error in decoy-based benchmarks and makes coming up with benchmarking sets easier.

Additionally, the \EFB{} has no dependence on the ratios of actives to random in the set, avoiding the problem where the EF maxes out at the overall ratio of actives. Instead, the \EFB{} achieves its maximum value at $\frac{1}{\chi}$ (the same maximum value achievable by the true enrichment). The minimum value of $\chi$ we could use the \EFB{} on is $\frac{1}{N_R}$, where $N_R$ is the number of random compounds (this would correspond to $S_\chi$ being the maximum score on the random set. This is a much more efficient use of the data than the EF, which requires $\frac{N_S}{\chi}$ total compounds to measure enrichment at $\chi$, where $N_S$ is the number of selected compounds (if $N_S$ is too small, the EF becomes noisy).

We additionally note that, as na\"ive ratio estimators, both the \EF{} and \EFB{} formulae are biased estimators of the true enrichment $\text{EF}^T$. This is an undesirable property, though we leave the creation of an unbiased estimator as future work. For now, we simply recommend paying close attention to the computed confidence intervals around these metrics.

\section{The maximum Bayes enrichment factor}

\begin{figure}
	\centering
	\includegraphics[width=0.8\linewidth]{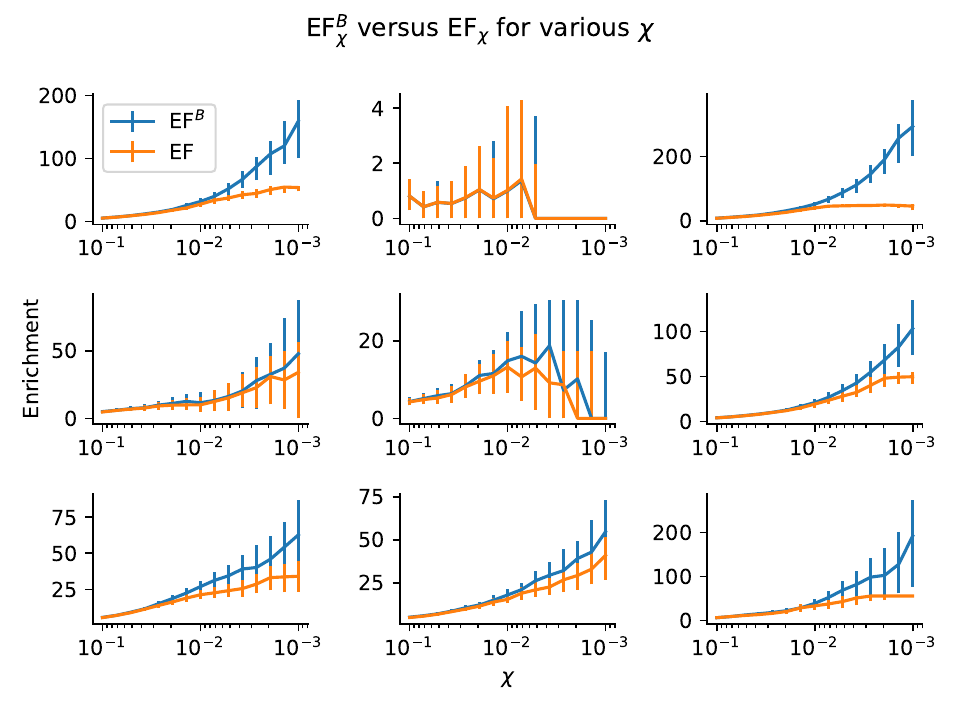}
	\caption{\EFB{} and EF (along with 95\% confidence intervals) for various $\chi$ for randomly selected models and targets on DUD-E (data from \citet{sunseri_virtual_2021}). Notice that the \EFB{} can be significantly higher than the EF at low $\chi$ due to the maxing-out problem of the EF. Both estimates can converge to 0 at sufficiently low $\chi$ due to both metrics failing to select any active molecules above the current threshold.}
	\label{fig:efb_chi}
\end{figure}

The \EFB{} allows us to estimate enrichments for $\chi$ as low as $\frac{1}{N_R}$, where $N_R$ is the number of random compounds in the set. This raises an important question: what value of $\chi$ should we measure enrichment at? As can be seen in Figure \ref{fig:efb_chi}, values at extremely low $\chi$ often have large confidence intervals and sometimes become 0 due to the lack of any known actives with scores above $S_\chi$.

We propose simply taking the maximum value of the \EFB{} achieved over the measurable $\chi$ interval of $[\frac{1}{N_R}, 1]$, which we denote the \EFB{max}. If we assume that the true enrichment $\text{EF}^T_\chi$ increases monotonically as $\chi \rightarrow 0$ (that is, that the probability of success in a virtual screen increases with the size of the screen), then the \EFB{max} is our best guess at how well the model will do in a real-life screen.

We do note that, because the \EFB{max} usually occurs at very low $\chi$, its confidence interval is often very wide. It is unfortunate that we cannot pin down the true value with great certainty, but valuable information can still be gleaned. We specifically suggest using the \textit{lower confidence bound} as an informative metric. This tells us, with high confidence, that our model will \textit{at least this well} in a prospective virtual screen for the current target.

\section{Metric comparison on DUD-E}

In order to compare \EF{} and \EFB{} metrics, we analyzed the performance of several models on the DUD-E benchmark. These models had been previously benchmarked in \citet{sunseri_virtual_2021}; we used the data from their study and re-analyzed it using both the \EF{} and \EFB{}. We report the median metrics across all DUD-E targets in Table \ref{tab:dude}.

From this table, we can make a couple of observations. First, the \EF{} and \EFB{} agree well at $\chi = 1\%$ (the most common \EF{} value reported). At $\chi = 0.1\%$, however, the \EFB{} results are much higher. This is due to the problem outlined above wherein the EF maxes out at the decoy to actives ratio. The \EFB{} suffers no such drawbacks, and can thus be used to more accurately assess model performance at low $\chi$.

\begin{table}[!hbt]
\centering
\caption{Comparison of EF and \EFB{} metrics for models on the DUD-E benchmarking set. Results were taken from \citet{sunseri_virtual_2021}.}
\label{tab:dude}
\resizebox{\textwidth}{!}{%
\begin{tabular}{llllll}
\toprule
Model & $\text{EF}_{1\%}$ & $\text{EF}^B_{1\%}$ & $\text{EF}_{0.1\%}$ & $\text{EF}^B_{0.1\%}$ & \EFB{max} \\
\midrule
Vina & 7.0 [6.6, 8.3] & 7.7 [7.1, 9.1] & 11 [7.2, 13] & 12 [7.8, 15] & 32 [21, 34] \\
Vinardo & 11 [9.8, 12] & 12 [11, 13] & 20. [14, 22] & 20. [17, 25] & 48 [36, 56] \\
General (Affinity) & 12 [10., 13] & 13 [11, 15] & 20. [17, 26] & 26 [21, 34] & 61 [43, 70.] \\
General (Pose) & 10. [8.7, 11] & 11 [9.4, 12] & 17 [11, 19] & 17 [13, 24] & 37 [32, 50.] \\
Dense (Affinity) & 18 [16, 18] & 21 [18, 22] & 32 [28, 38] & 55 [42, 57] & 100 [81, 120] \\
Dense (Pose) & 21 [18, 22] & 23 [21, 25] & 42 [37, 45] & 77 [59, 84] & 160 [130, 180] \\
Default (Affinity) & 15 [14, 17] & 18 [16, 19] & 32 [28, 38] & 50. [39, 56] & 130 [87, 140] \\
Default (Pose) & 16 [14, 17] & 19 [16, 19] & 32 [28, 37] & 45 [37, 53] & 130 [81, 130] \\
\bottomrule

\end{tabular}
}
\end{table}

\section{The BayesBind benchmark}

The fact that the \EFB{} requires random compounds rather than true inactives enables much simpler benchmark creation. We sought to capitalize on this by creating the BayesBind benchmark, which we also explicitly designed for easier benchmarking of ML-based models.

The BayesBind benchmark is composed of validation and test sets whose targets are taken from the validation and test sets of the BigBind set, respectively. Molecules with known activity for each target were clustered according to a Tanimoto similarity cutoff of 0.4 on 2048-bit Morgan fingerprints with radius 3. Targets were chosen with at least 50 clusters containing molecules with a pChEMBL value greater than 5 (10 \textmu M). To limit the redundancy in the set, we chose a single molecule from each cluster (the molecule with median pChEMBL value). We saved all such molecules in our ``actives'' set, even those below 10 \textmu M; this is so that users can choose different activity cutoff thresholds when analyzing their results.

To ensure the diversity of the targets of the benchmark, only one target was chosen for every pocket cluster (as defined by the pocket-level TM-score clusters used to split the BigBind set). We chose the target in each pocket cluster that would yield the highest number of actives for the benchmark. 

For each target, a set of 1,000 random compounds was generated. These compounds were chosen from the set of compounds in the BigBind validation and test sets not known to bind to any pockets in the same pocket cluster as the current target. The compounds were clustered according to the same Tanimoto cutoff of 0.4, and a random compound was chosen from each cluster. We note that this selection of the random set guarantees that the active and random molecules come from the same region of chemical space. This is not the case in DUD-E, where decoys are generated from ZINC \cite{irwin_zinc_2005} and actives from ChEMBL \cite{gaulton_chembl_2012}. Even with the property matching employed by DUD-E, subtle differences still exist between the sets that can be exploited by ML models \cite{chen_hidden_2019, yang_predicting_2020}.

The final BayesBind set contains 14 targets in the validation set and 11 in the test set. There are an average of 297 known activities for each target, and an average of 187 actives with activity less than 10 \textmu M.

\section{The BayesBind ML subset}

We desire to use the BayesBind set to rigorously benchmark ML models trained on the BigBind training set. Therefore, we desire a simple baseline with which to compare the performance of more complex models. To validate the splits of the BigBind set, we used a K-nearest-neighbors (KNN) classifier based on the ligand and pocket similarities of protein-ligand pairs in the train set \cite{brocidiacono_bigbind_2023}. The KNN did no better than random on the BigBind test set; we used this as evidence that our splitting was rigorous.

We used this same KNN to validate the BayesBind benchmarking set. We were surprised to discover that the KNN significantly outperformed all other models, including a neural network model (\textsc{Banana}) that was also trained on BigBind \cite{brocidiacono_bigbind_2023}. On some targets, it was able to achieve enormous enrichment values (\EFB{max} up to 233). The median \EFB{max} it achieved on the full BayesBind set was 34 [21, 51] on the validation set and 11 [4.8, 21] on the test set.

How do we reconcile the KNN's performance on the BayesBind set with its poor classification performance on the BigBind test set overall? We argue that the BigBind splits were mostly successful at placing too-similar pockets in the same split, but had some noise. The BayesBind set is only composed of targets with many known active molecules, and therefore are likely to be in well-studied protein families. Thus there are likely to be many similar protein-ligand pairs across the BigBind set as a whole; because of the noise on the similarity metrics used, some of these were placed in the training set. It only takes a single similar datapoint in the training set to massively increase the performance of the KNN.

The performance of the KNN on this set is concerning and more research should be done on even more rigorous data splits. For the time being, however, we decided to create an ML subset of the BayesBind set composed only of targets that the KNN did relatively poorly on (\EFB{max} less than 30). While the KNN still achieves nonrandom enrichment on many targets in this subset (and in fact still outperforms all the other models), there is still much room for improvement. A compelling ML model should be able to significantly outperform the KNN on this benchmark.

\section{Benchmarking results}

We evaluated three popular SBVS methods (AutoDock Vina \cite{trott_autodock_2010}, \textsc{Gnina} \cite{mcnutt_gnina_2021}, and Glide \cite{friesner_glide_2004}) on the benchmark, in addition to \textsc{Banana} and the KNN outlined above.

\begin{figure}[!htb]
	\centering
	\includegraphics[width=1.0\linewidth]{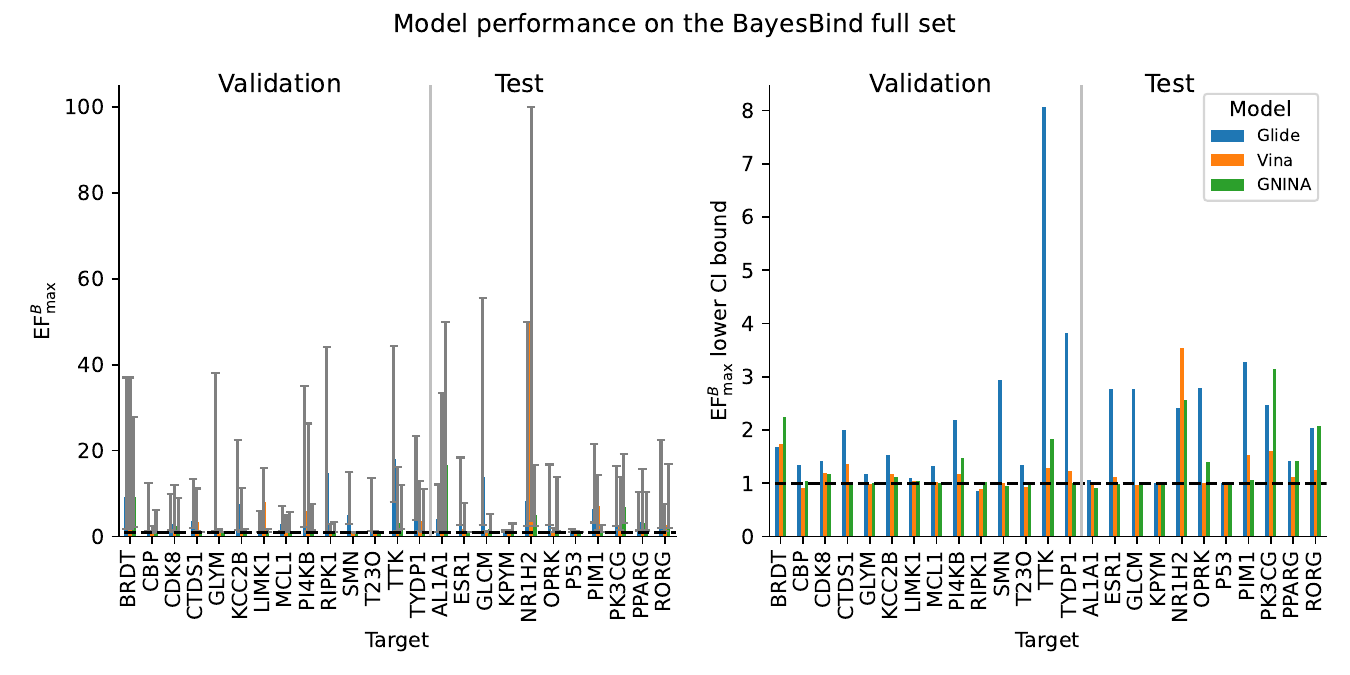}
	\caption{Left: \EFB{max} for Glide, Vina, and \textsc{Gnina} on each target in the full BayesBind set. Right: the \EFB{max} 95\% CI lower bounds for the same models and targets. On both plots, the dotted line indicates no enrichment (\EFB{} of 1).}
	\label{fig:baselines_full_max}
\end{figure}

\begin{figure}[!htb]
	\centering
	\includegraphics[width=1.0\linewidth]{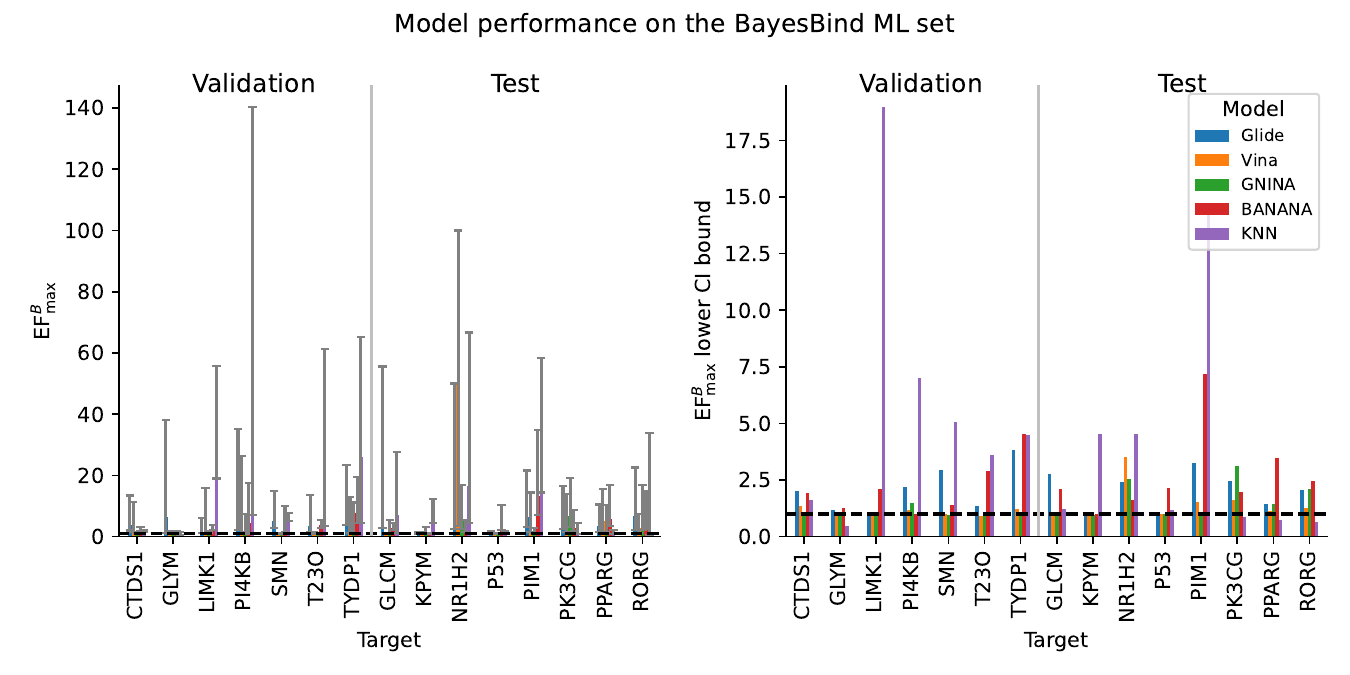}
	\caption{Left: \EFB{max} for all the models (including \textsc{Banana} and the KNN) on each target in the ML subset. Right: the \EFB{max} 95\% CI lower bounds for the same models and targets. On both plots, the dotted line indicates no enrichment (\EFB{} of 1).}
	\label{fig:baselines_ml_max}
\end{figure}

The results for the three models not trained on BigBind (Glide, Vina, and \textsc{Gnina}) are shown in figure \ref{fig:baselines_full_max}, and the median results across all targets are shown in Table \ref{tab:full_summary}. Similarly, the results for all the models (including \textsc{Banana} and the KNN) are shown in Figure \ref{fig:baselines_ml_max} and Table \ref{tab:ml_summary}.

Overall, we see a wide range of model performances; most models (except for the KNN) perform similarly to random for most targets, but some are able to achieve 
notable enrichment. We note, however, that even the highest \EFB{max} achieved (50. [3.5, 100] by Vina on NR1H2), is significantly less than the enrichments observed by models on DUD-E (see Table \ref{tab:dude} and Figure \ref{fig:efb_chi}). Indeed, it is much less than the target enrichment we desire for virtual screening to be competitive with HTS (1,000).

\begin{table}[!htb]
\centering
\caption{Median AUCs, \EFB{1\%}, and \EFB{max} results for Glide, \textsc{Gnina}, and Vina on the full BayesBind set. 95\% confidence intervals are shown in brackets.}
\label{tab:full_summary}
\resizebox{\textwidth}{!}{%
\begin{tabular}{lllllllll}
\toprule
\multirow{2.5}{*}{Model} & \multicolumn{3}{c}{Validation} & \multicolumn{3}{c}{Test}  \\
\cmidrule(lr){2-4} \cmidrule(lr){5-7}
 & AUC & \EFB{1\%} & \EFB{max} & AUC & \EFB{1\%} & \EFB{max} \\
\midrule
Glide & 0.57 [0.54, 0.59] & 2.0 [0.96, 3.0] & 5.7 [3.0, 8.3] & 0.50 [0.49, 0.56] & 3.6 [1.4, 4.2] & 5.5 [3.3, 8.3] \\
\textsc{Gnina} & 0.48 [0.44, 0.50] & 0.51 [0.0, 0.67] & 1.3 [1.1, 2.1] & 0.46 [0.43, 0.50] & 1.5 [0.0, 2.2] & 3.1 [1.2, 5.2] \\
Vina & 0.46 [0.45, 0.50] & 1.5 [0.058, 1.5] & 3.0 [1.3, 3.5] & 0.51 [0.43, 0.53] & 1.3 [0.21, 1.7] & 2.7 [1.1, 3.9] \\
\bottomrule
\end{tabular}
}
\end{table}

\begin{table}[!htb]
\centering
\caption{Median AUCs, \EFB{1\%}, and \EFB{max} results for all models (including \textsc{Banana} and the KNN) on the BayesBind ML subset. 95\% confidence intervals are shown in brackets.}
\label{tab:ml_summary}
\resizebox{\textwidth}{!}{%
\begin{tabular}{lllllllll}
\toprule
\multirow{2.5}{*}{Model} & \multicolumn{3}{c}{Validation} & \multicolumn{3}{c}{Test}  \\
\cmidrule(lr){2-4} \cmidrule(lr){5-7}
 & AUC & \EFB{1\%} & \EFB{max} & AUC & \EFB{1\%} & \EFB{max} \\
\midrule
KNN & 0.70 [0.64, 0.75] & 6.2 [0.12, 7.3] & 20. [5.1, 26] & 0.50 [0.48, 0.53] & 1.5 [0.43, 3.5] & 7.0 [1.5, 10.] \\
\textsc{Banana} & 0.72 [0.69, 0.74] & 0.20 [0.0, 1.1] & 2.5 [2.1, 4.3] & 0.66 [0.62, 0.68] & 2.6 [0.68, 4.0] & 3.1 [2.7, 5.9] \\
Glide & 0.57 [0.53, 0.59] & 2.3 [0.72, 3.9] & 5.0 [2.7, 8.7] & 0.52 [0.50, 0.57] & 3.4 [1.2, 4.0] & 5.9 [2.9, 9.6] \\
\textsc{Gnina} & 0.37 [0.34, 0.40] & 0.0 [0.0, 0.0] & 1.0 [1.0, 1.1] & 0.54 [0.51, 0.56] & 1.6 [0.14, 2.8] & 2.5 [1.5, 4.7] \\
Vina & 0.45 [0.40, 0.49] & 1.8 [0.0, 2.2] & 3.4 [1.1, 4.5] & 0.52 [0.46, 0.54] & 1.5 [0.23, 2.1] & 2.7 [1.3, 4.7] \\
\bottomrule
\end{tabular}
}
\end{table}

\section{Conclusion}

We have shown that the traditional formula for estimating virtual screening performance, the enrichment factor, fails to estimate the true enrichment of a model in a real-life setting (very low active-to-inactive ratio and low selection fraction $\chi$). We propose a new formula, the \EFB{}, that is just as easy to calculate but is able to better estimate enrichment in realistic scenarios. Notably, this new formula only requires a set of known active molecules and a set of random molecules. We also suggest the use of the maximum Bayes enrichment factor (\EFB{max}), and especially its lower confidence bound, as a key summary metric for a model on a VS benchmark.

We also developed a new benchmarking set, BayesBind, that pairs known actives for targets from the BigBind validation and test sets with random compounds. This set enables easy benchmarking of ML models that are trained on BigBind. We evaluated three common VS models on this benchmark, along with KNN and a neural network model (\textsc{Banana}). 

Our analysis shows that the VS models studied generally fail to achieve nonrandom enrichment on the targets in the set. This failure stands in contrast to the reported success of these methods on decoy-based benchmarks such as DUD-E. We hypothesize two possible reasons for this.

First, the ``active'' modules for DUD-E targets tend to be tighter binders than those in the BayesBind set; DUD-E selects the lowest-affinity ligands in clusters and has reduced affinity cutoff values (down to 3 nM) for targets with many known actives. BayesBind, in contrast, takes the median-affinity ligand from each cluster. All our analysis here uses a 10 \textmu M activity cutoff. We argue that this higher cutoff is much more realistic; nanomolar binders are vanishingly rare and almost never show up in primary screens. For now, we should pursue models that can reliably discover low \textmu M binders.

Second, the choice of decoys in DUD-E could have a large effect on model performance. DUD-E chooses decoys from ZINC and its actives from ChEMBL; even though molecular properties are matched across the sets, there are still subtle differences that, for instance, ML models can abuse \cite{chen_hidden_2019}. BayesBind, by contrast, uses random compounds (not decoys) that are taken from the same chemical space (BigBind compounds, which ultimately come from ChEMBL). Notably, all the compounds in BigBind are annotated with activity against at least one target; the set could be enriched in nonspecific protein binders. If this is the case, the BayesBind set is also implicitly testing the model's ability to find \textit{selective} compounds to some extent. This is a harder task, and an important one; real-life drug discovery projects are generally only interested in selective binders.

Overall, the BayesBind set is an important new benchmark for estimating VS model performance in a real-life scenario. The fact that no model does well on it stands as a testament to the difficulty of prospective virtual screening. We hope that our comrades will use this benchmark to rise to the challenge of making new methods succeed in this area. 

\section{Acknowledgements}
The authors thank David Koes, James Wellnitz, Travis Maxfield, Nyssa Tucker, and Enes Kelestemur for their support and insight as we worked out the details of this study. Studies described in this paper were supported by the NIH grant R01GM140154.

\bibliographystyle{unsrtnat}
\bibliography{references}

\begin{thebibliography}{18}
\providecommand{\natexlab}[1]{#1}
\providecommand{\url}[1]{\texttt{#1}}
\expandafter\ifx\csname urlstyle\endcsname\relax
  \providecommand{\doi}[1]{doi: #1}\else
  \providecommand{\doi}{doi: \begingroup \urlstyle{rm}\Url}\fi

\bibitem[Mysinger et~al.(2012)Mysinger, Carchia, Irwin, and Shoichet]{mysinger_directory_2012}
Michael~M. Mysinger, Michael Carchia, John.~J. Irwin, and Brian~K. Shoichet.
\newblock Directory of {Useful} {Decoys}, {Enhanced} ({DUD}-{E}): {Better} {Ligands} and {Decoys} for {Better} {Benchmarking}.
\newblock \emph{Journal of Medicinal Chemistry}, 55\penalty0 (14):\penalty0 6582--6594, July 2012.
\newblock ISSN 0022-2623.
\newblock \doi{10.1021/jm300687e}.
\newblock URL \url{https://doi.org/10.1021/jm300687e}.
\newblock Publisher: American Chemical Society.

\bibitem[Su et~al.(2019)Su, Yang, Du, Feng, Liu, Li, and Wang]{su_comparative_2019}
Minyi Su, Qifan Yang, Yu~Du, Guoqin Feng, Zhihai Liu, Yan Li, and Renxiao Wang.
\newblock Comparative {Assessment} of {Scoring} {Functions}: {The} {CASF}-2016 {Update}.
\newblock \emph{Journal of Chemical Information and Modeling}, 59\penalty0 (2):\penalty0 895--913, February 2019.
\newblock ISSN 1549-9596.
\newblock \doi{10.1021/acs.jcim.8b00545}.
\newblock URL \url{https://doi.org/10.1021/acs.jcim.8b00545}.
\newblock Publisher: American Chemical Society.

\bibitem[Tran-Nguyen et~al.(2020)Tran-Nguyen, Jacquemard, and Rognan]{tran-nguyen_lit-pcba_2020}
Viet-Khoa Tran-Nguyen, Célien Jacquemard, and Didier Rognan.
\newblock {LIT}-{PCBA}: {An} {Unbiased} {Data} {Set} for {Machine} {Learning} and {Virtual} {Screening}.
\newblock \emph{Journal of Chemical Information and Modeling}, 60\penalty0 (9):\penalty0 4263--4273, September 2020.
\newblock ISSN 1549-9596.
\newblock \doi{10.1021/acs.jcim.0c00155}.
\newblock URL \url{https://doi.org/10.1021/acs.jcim.0c00155}.
\newblock Publisher: American Chemical Society.

\bibitem[Wellnitz et~al.(2023)Wellnitz, Jain, Hochuli, Maxfield, Muratov, Tropsha, and Zakharov]{wellnitz_one_2023}
James Wellnitz, Sankalp Jain, Joshua Hochuli, Travis Maxfield, Eugene Muratov, Alexander Tropsha, and Alexey Zakharov.
\newblock One size does not fit all: revising traditional paradigms for {QSAR}-based virtual screenings., December 2023.
\newblock URL \url{https://chemrxiv.org/engage/chemrxiv/article-details/6585ddc19138d23161476eb1}.

\bibitem[Yang et~al.(2020)Yang, Shen, and Huang]{yang_predicting_2020}
Jincai Yang, Cheng Shen, and Niu Huang.
\newblock Predicting or {Pretending}: {Artificial} {Intelligence} for {Protein}-{Ligand} {Interactions} {Lack} of {Sufficiently} {Large} and {Unbiased} {Datasets}.
\newblock \emph{Frontiers in Pharmacology}, 11, 2020.
\newblock ISSN 1663-9812.
\newblock URL \url{https://www.frontiersin.org/journals/pharmacology/articles/10.3389/fphar.2020.00069}.

\bibitem[Francoeur et~al.(2020)Francoeur, Masuda, Sunseri, Jia, Iovanisci, Snyder, and Koes]{francoeur_three-dimensional_2020}
Paul~G. Francoeur, Tomohide Masuda, Jocelyn Sunseri, Andrew Jia, Richard~B. Iovanisci, Ian Snyder, and David~R. Koes.
\newblock Three-{Dimensional} {Convolutional} {Neural} {Networks} and a {Cross}-{Docked} {Data} {Set} for {Structure}-{Based} {Drug} {Design}.
\newblock \emph{Journal of Chemical Information and Modeling}, 60\penalty0 (9):\penalty0 4200--4215, September 2020.
\newblock ISSN 1549-9596.
\newblock \doi{10.1021/acs.jcim.0c00411}.
\newblock URL \url{https://doi.org/10.1021/acs.jcim.0c00411}.
\newblock Publisher: American Chemical Society.

\bibitem[Volkov et~al.(2022)Volkov, Turk, Drizard, Martin, Hoffmann, Gaston-Mathé, and Rognan]{volkov_frustration_2022}
Mikhail Volkov, Joseph-André Turk, Nicolas Drizard, Nicolas Martin, Brice Hoffmann, Yann Gaston-Mathé, and Didier Rognan.
\newblock On the {Frustration} to {Predict} {Binding} {Affinities} from {Protein}–{Ligand} {Structures} with {Deep} {Neural} {Networks}.
\newblock \emph{Journal of Medicinal Chemistry}, 65\penalty0 (11):\penalty0 7946--7958, June 2022.
\newblock ISSN 0022-2623.
\newblock \doi{10.1021/acs.jmedchem.2c00487}.
\newblock URL \url{https://doi.org/10.1021/acs.jmedchem.2c00487}.
\newblock Publisher: American Chemical Society.

\bibitem[Brocidiacono et~al.(2023)Brocidiacono, Francoeur, Aggarwal, Popov, Koes, and Tropsha]{brocidiacono_bigbind_2023}
Michael Brocidiacono, Paul Francoeur, Rishal Aggarwal, Konstantin~I. Popov, David~Ryan Koes, and Alexander Tropsha.
\newblock {BigBind}: {Learning} from {Nonstructural} {Data} for {Structure}-{Based} {Virtual} {Screening}.
\newblock \emph{Journal of Chemical Information and Modeling}, December 2023.
\newblock ISSN 1549-9596.
\newblock \doi{10.1021/acs.jcim.3c01211}.
\newblock URL \url{https://doi.org/10.1021/acs.jcim.3c01211}.
\newblock Publisher: American Chemical Society.

\bibitem[Sunseri and Koes(2021)]{sunseri_virtual_2021}
Jocelyn Sunseri and David~Ryan Koes.
\newblock Virtual {Screening} with {Gnina} 1.0.
\newblock \emph{Molecules}, 26\penalty0 (23):\penalty0 7369, January 2021.
\newblock ISSN 1420-3049.
\newblock \doi{10.3390/molecules26237369}.
\newblock URL \url{https://www.mdpi.com/1420-3049/26/23/7369}.
\newblock Number: 23 Publisher: Multidisciplinary Digital Publishing Institute.

\bibitem[Irwin and Shoichet(2005)]{irwin_zinc_2005}
John~J. Irwin and Brian~K. Shoichet.
\newblock {ZINC} – {A} {Free} {Database} of {Commercially} {Available} {Compounds} for {Virtual} {Screening}.
\newblock \emph{Journal of chemical information and modeling}, 45\penalty0 (1):\penalty0 177--182, 2005.
\newblock ISSN 1549-9596.
\newblock \doi{10.1021/ci049714}.
\newblock URL \url{https://www.ncbi.nlm.nih.gov/pmc/articles/PMC1360656/}.

\bibitem[Gaulton et~al.(2012)Gaulton, Bellis, Bento, Chambers, Davies, Hersey, Light, McGlinchey, Michalovich, Al-Lazikani, and Overington]{gaulton_chembl_2012}
Anna Gaulton, Louisa~J. Bellis, A.~Patricia Bento, Jon Chambers, Mark Davies, Anne Hersey, Yvonne Light, Shaun McGlinchey, David Michalovich, Bissan Al-Lazikani, and John~P. Overington.
\newblock {ChEMBL}: a large-scale bioactivity database for drug discovery.
\newblock \emph{Nucleic Acids Research}, 40\penalty0 (Database issue):\penalty0 D1100--D1107, January 2012.
\newblock ISSN 0305-1048.
\newblock \doi{10.1093/nar/gkr777}.
\newblock URL \url{https://www.ncbi.nlm.nih.gov/pmc/articles/PMC3245175/}.

\bibitem[Chen et~al.(2019)Chen, Cruz, Ramsey, Dickson, Duca, Hornak, Koes, and Kurtzman]{chen_hidden_2019}
Lieyang Chen, Anthony Cruz, Steven Ramsey, Callum~J. Dickson, Jose~S. Duca, Viktor Hornak, David~R. Koes, and Tom Kurtzman.
\newblock Hidden bias in the {DUD}-{E} dataset leads to misleading performance of deep learning in structure-based virtual screening.
\newblock \emph{PLOS ONE}, 14\penalty0 (8):\penalty0 e0220113, August 2019.
\newblock ISSN 1932-6203.
\newblock \doi{10.1371/journal.pone.0220113}.
\newblock URL \url{https://journals.plos.org/plosone/article?id=10.1371/journal.pone.0220113}.
\newblock Publisher: Public Library of Science.

\bibitem[Trott and Olson(2010)]{trott_autodock_2010}
Oleg Trott and Arthur~J. Olson.
\newblock {AutoDock} {Vina}: improving the speed and accuracy of docking with a new scoring function, efficient optimization and multithreading.
\newblock \emph{Journal of computational chemistry}, 31\penalty0 (2):\penalty0 455--461, January 2010.
\newblock ISSN 0192-8651.
\newblock \doi{10.1002/jcc.21334}.
\newblock URL \url{https://www.ncbi.nlm.nih.gov/pmc/articles/PMC3041641/}.

\bibitem[McNutt et~al.(2021)McNutt, Francoeur, Aggarwal, Masuda, Meli, Ragoza, Sunseri, and Koes]{mcnutt_gnina_2021}
Andrew~T. McNutt, Paul Francoeur, Rishal Aggarwal, Tomohide Masuda, Rocco Meli, Matthew Ragoza, Jocelyn Sunseri, and David~Ryan Koes.
\newblock {GNINA} 1.0: molecular docking with deep learning.
\newblock \emph{Journal of Cheminformatics}, 13\penalty0 (1):\penalty0 43, June 2021.
\newblock ISSN 1758-2946.
\newblock \doi{10.1186/s13321-021-00522-2}.
\newblock URL \url{https://doi.org/10.1186/s13321-021-00522-2}.

\bibitem[Friesner et~al.(2004)Friesner, Banks, Murphy, Halgren, Klicic, Mainz, Repasky, Knoll, Shelley, Perry, Shaw, Francis, and Shenkin]{friesner_glide_2004}
Richard~A. Friesner, Jay~L. Banks, Robert~B. Murphy, Thomas~A. Halgren, Jasna~J. Klicic, Daniel~T. Mainz, Matthew~P. Repasky, Eric~H. Knoll, Mee Shelley, Jason~K. Perry, David~E. Shaw, Perry Francis, and Peter~S. Shenkin.
\newblock Glide: {A} {New} {Approach} for {Rapid}, {Accurate} {Docking} and {Scoring}.
\newblock \emph{Journal of Medicinal Chemistry}, 47\penalty0 (7):\penalty0 1739--1749, March 2004.
\newblock ISSN 0022-2623.
\newblock \doi{10.1021/jm0306430}.
\newblock URL \url{https://doi.org/10.1021/jm0306430}.
\newblock Publisher: American Chemical Society.

\bibitem[Virtanen et~al.(2020)Virtanen, Gommers, Oliphant, Haberland, Reddy, Cournapeau, Burovski, Peterson, Weckesser, Bright, van~der Walt, Brett, Wilson, Millman, Mayorov, Nelson, Jones, Kern, Larson, Carey, Polat, Feng, Moore, VanderPlas, Laxalde, Perktold, Cimrman, Henriksen, Quintero, Harris, Archibald, Ribeiro, Pedregosa, and van Mulbregt]{virtanen_scipy_2020}
Pauli Virtanen, Ralf Gommers, Travis~E. Oliphant, Matt Haberland, Tyler Reddy, David Cournapeau, Evgeni Burovski, Pearu Peterson, Warren Weckesser, Jonathan Bright, Stéfan~J. van~der Walt, Matthew Brett, Joshua Wilson, K.~Jarrod Millman, Nikolay Mayorov, Andrew R.~J. Nelson, Eric Jones, Robert Kern, Eric Larson, C.~J. Carey, İlhan Polat, Yu~Feng, Eric~W. Moore, Jake VanderPlas, Denis Laxalde, Josef Perktold, Robert Cimrman, Ian Henriksen, E.~A. Quintero, Charles~R. Harris, Anne~M. Archibald, Antônio~H. Ribeiro, Fabian Pedregosa, and Paul van Mulbregt.
\newblock {SciPy} 1.0: fundamental algorithms for scientific computing in {Python}.
\newblock \emph{Nature Methods}, 17\penalty0 (3):\penalty0 261--272, March 2020.
\newblock ISSN 1548-7105.
\newblock \doi{10.1038/s41592-019-0686-2}.
\newblock URL \url{https://www.nature.com/articles/s41592-019-0686-2}.
\newblock Publisher: Nature Publishing Group.

\bibitem[O'Boyle et~al.(2011)O'Boyle, Banck, James, Morley, Vandermeersch, and Hutchison]{oboyle_open_2011}
Noel~M. O'Boyle, Michael Banck, Craig~A. James, Chris Morley, Tim Vandermeersch, and Geoffrey~R. Hutchison.
\newblock Open {Babel}: {An} open chemical toolbox.
\newblock \emph{Journal of Cheminformatics}, 3\penalty0 (1):\penalty0 33, October 2011.
\newblock ISSN 1758-2946.
\newblock \doi{10.1186/1758-2946-3-33}.
\newblock URL \url{https://doi.org/10.1186/1758-2946-3-33}.

\bibitem[noa(2024)]{noauthor_forlilabmeeko_2024}
forlilab/{Meeko}, March 2024.
\newblock URL \url{https://github.com/forlilab/Meeko}.
\newblock original-date: 2020-11-07T12:05:36Z.

\end{thebibliography}

\appendix
\section{Appendix}

\makeatletter
\renewcommand\thetable{A.\@arabic\c@table}
\renewcommand \thefigure{A.\@arabic\c@figure}
\makeatother

\subsection{Computing confidence intervals}

All confidence intervals reported in this study were computed using SciPy's \texttt{stats.bootstrap} method with 1,000 samples and the ``percentile'' method \cite{virtanen_scipy_2020}.

\subsection{Benchmarking Glide}

Before running Glide, we prepared the ligands for docking with \texttt{ligprep -i 2}. Similarly, we prepared each receptor with \texttt{prepwizard -fix}. When docking, we used the box sizes specified from the BigBind dataset. Glide also requires an inner box boz size within which to keep the ligand centroid; we used half the outer box size. We used the \texttt{r\_i\_docking\_score} as the score for each ligand.

\subsection{Benchmarking AutoDock Vina}
All receptors were converted to PDBQT files using Open Babel \cite{oboyle_open_2011} and ligands were converted with Meeko \cite{noauthor_forlilabmeeko_2024}. Vina was run with the BigBind box sizes.

\subsection{Benchmarking \textsc{Gnina}}
\textsc{Gnina} was also run with the BigBind box sizes, and with the additional command line arguments \texttt{--cnn crossdock\_default2018}. We used the \texttt{CNNaffinity} output as the score for each ligand.

\subsection{Full benchmarking results}

\begin{figure}[!htb]
	\centering
	\includegraphics[width=1.0\linewidth]{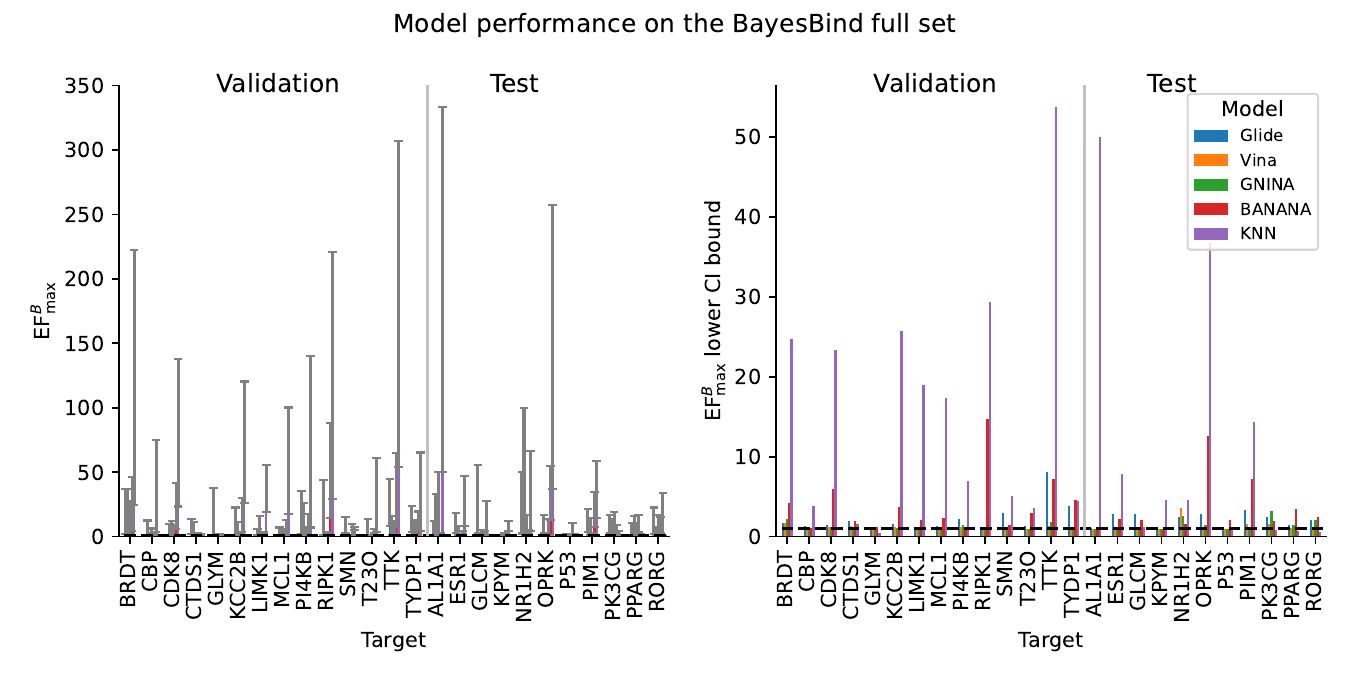}
	\caption{Left: \EFB{max} for every model (including the ML models KNN and \textsc{Banana}) on each target in the full BayesBind set. Right: the \EFB{max} lower bounds for the same models and targets. On both plots, the dotted line indicates no enrichment (\EFB{} of 1).}
	\label{fig:baselines_full_full_max}
\end{figure}

\begin{table}[!htb]
\centering
\caption{Median results of all models on the full BayesBind set}
\label{tab:ml_full_summary}
\resizebox{\textwidth}{!}{%
\begin{tabular}{lllllllll}
\toprule
\multirow{2.5}{*}{Model} & \multicolumn{3}{c}{Validation} & \multicolumn{3}{c}{Test}  \\
\cmidrule(lr){2-4} \cmidrule(lr){5-7}
 & AUC & \EFB{1\%} & \EFB{max} & AUC & \EFB{1\%} & \EFB{max} \\
\midrule
KNN & 0.68 [0.65, 0.72] & 8.0 [5.5, 11] & 34 [21, 51] & 0.55 [0.50, 0.57] & 7.0 [2.7, 8.0] & 11 [4.8, 21] \\
\textsc{Banana} & 0.69 [0.67, 0.72] & 2.3 [0.73, 3.8] & 4.3 [3.3, 8.1] & 0.67 [0.63, 0.69] & 2.5 [0.40, 3.5] & 3.0 [2.6, 5.5] \\
Glide & 0.57 [0.54, 0.59] & 2.0 [0.91, 3.0] & 5.7 [2.9, 8.0] & 0.50 [0.49, 0.56] & 3.6 [1.4, 4.3] & 5.5 [3.3, 8.2] \\
\textsc{Gnina} & 0.48 [0.44, 0.50] & 0.51 [0.0, 0.58] & 1.3 [1.2, 2.1] & 0.46 [0.43, 0.50] & 1.5 [0.0, 2.5] & 3.1 [1.2, 5.2] \\
Vina & 0.46 [0.45, 0.50] & 1.5 [0.15, 1.5] & 3.0 [1.4, 3.5] & 0.51 [0.44, 0.53] & 1.3 [0.31, 1.7] & 2.7 [1.1, 3.9] \\
\bottomrule
\end{tabular}
}
\end{table}

\end{document}